\documentclass[conference,10pt]{IEEEtran}
\IEEEoverridecommandlockouts
\usepackage{graphicx}
\graphicspath{ {./images/} }
\usepackage{amsmath}
\usepackage{comment}
\usepackage{dsfont}
\usepackage{tabularx}
\usepackage{hyperref}
\usepackage{amsmath,amssymb,amsfonts}
\usepackage{algorithmic}
\usepackage{graphicx}
\usepackage{textcomp}
\usepackage{xcolor}
\usepackage{mathtools}
\usepackage{placeins}
\usepackage[ruled]{algorithm2e}
  
\usepackage{graphicx}
\usepackage{subcaption}
\usepackage[export]{adjustbox}
\usepackage{algorithmic}
\usepackage{algorithm2e}
\usepackage{wrapfig}
\usepackage{setspace}
\usepackage[font=small]{caption}
\begin{document}

\title{\huge{Semantic-Aware and Goal-Oriented Communications for Object Detection in Wireless  End-to-End Image
Transmission}}
\author{
   \IEEEauthorblockN{Fatemeh Zahra Safaeipour, Morteza Hashemi}
     \IEEEauthorblockA{Department of Electrical Engineering and Computer Science, University of Kansas }
\IEEEauthorblockA{Emails: \{fzsafaei, mhashemi\}@ku.edu}
}

\maketitle

\begin{abstract}
Semantic communication is focused on optimizing the exchange of information by transmitting only the most relevant data required to convey the intended message to the receiver and achieve the desired communication goal. For example, if we consider images as the information and the goal of the communication is object detection at the receiver side, the semantic of information would be the objects in each image. Therefore, by only transferring the semantics of images we can achieve the communication goal.
In this paper, we propose a design framework for implementing semantic-aware and goal-oriented communication of images. To achieve this, we first define the baseline problem as a set of mathematical problems that can be optimized to improve the efficiency and effectiveness of the communication system. We consider two scenarios in which either the data rate or the error at the receiver is the limiting constraint. Our proposed system model and solution is inspired by the concept of auto-encoders, where the encoder and the decoder are respectively implemented at the transmitter and receiver to extract semantic information for specific object detection goals. Our numerical results validate the proposed design framework to achieve low error or near-optimal in a goal-oriented communication system while reducing the amount of data transfers.

\end{abstract}


\section{Introduction}
 Shannon and Weaver in \cite{shannon2001mathematical} and \cite{Weaver2009RecentCT} proposed three levels for organizing the broad topic of communication: \textbf{(i) technical problem: } How precisely can communication symbols be transmitted?  \textbf{(ii) semantic problem: } 
How accurately do the symbols being transmitted convey the intended meaning? \textbf{(iii) effectiveness problem: } 
How effectively does the intended meaning influence the behavior of the receiver? Shannon established the foundation for information theory with his exact and formal solution to the technical problem. 
More recently, there has been an increasing interest to study the semantic and effectiveness problems.  The core of semantic communication is to extract the ``meanings'' of sent information at a transmitter and successfully ``interpret'' the semantic information at a receiver using a matched knowledge base between a transmitter and a receiver\cite{9679803}. Therefore,  semantic communication methods are able to identify relevant information that is strictly required to recover the meaning intended by the transmitter or to achieve a \emph{goal}, thereby enabling \emph{goal-oriented communication} and improving the overall data rate efficiency by using computing resources.
For example, if we aim to communicate images and we know that the purpose at the receiver side is to detect objects within each image, we can optimize our communication system to fulfill that goal, 
which is feasible with sufficient computing resources at both the transmitter and the receiver sides.

\begin{figure}[t]
    \centering
    \includegraphics[width=0.49\textwidth]{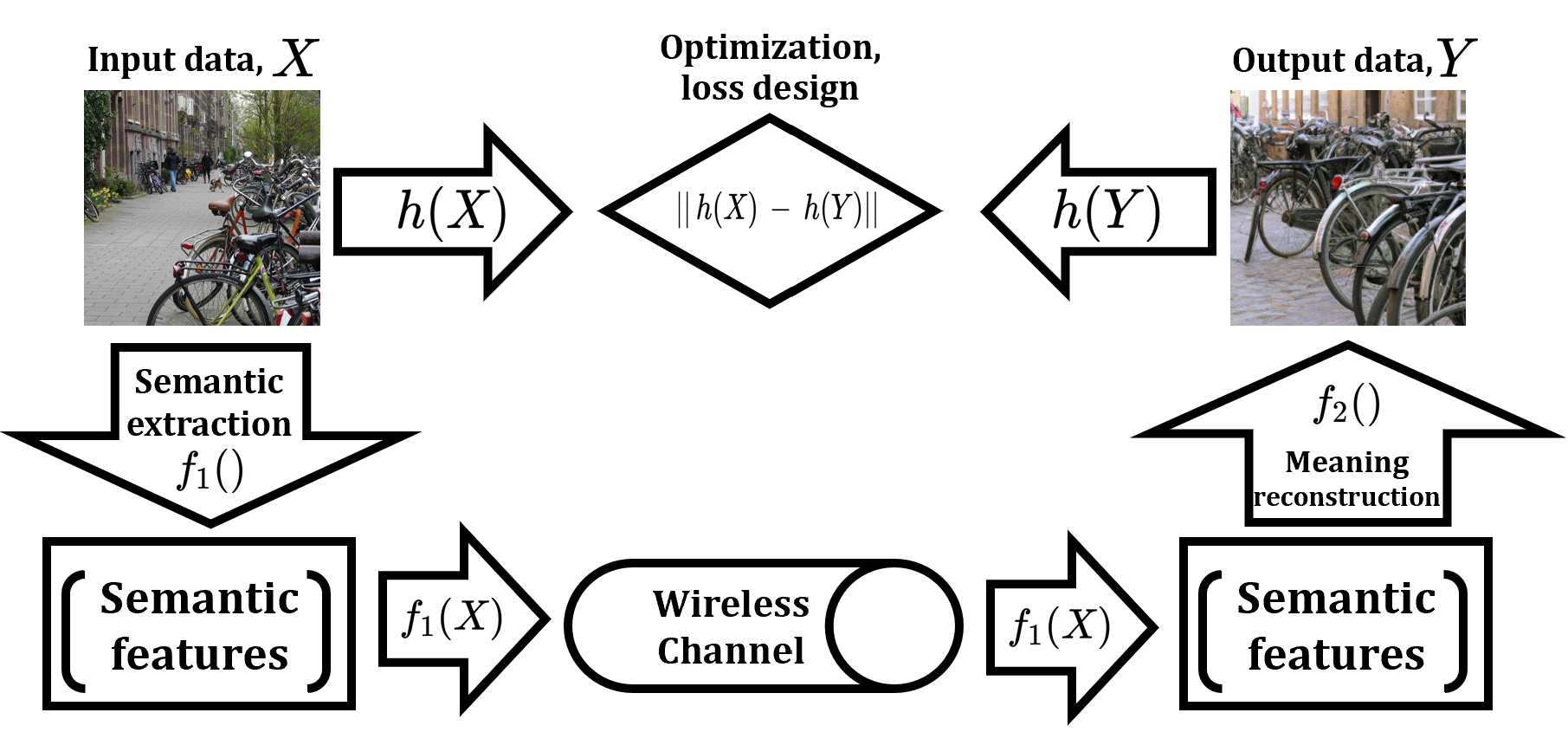}
    \caption{{System model for semantic-aware communication wherein the communication goal is to detect objects in images at the receiver. Function $f_1$ extracts semantic features, while function $f_2$ reconstructs the image from semantic features. Function $h$ measures the performance with respect to the communication goal. } }
    \label{fig:sysmodel}
\end{figure}
In this context, focusing on semantics and clearly defining the goal of communication assist us in distilling the data that are strictly relevant to fulfilling a predefined goal. Ignoring irrelevant data becomes a key strategy for significantly reducing the amount of data that must be transmitted and recovered, saving bandwidth, delay, and energy. 
Due to such promising performance gains compared with traditional communication systems, there has been an increasing interest to define semantic-aware and goal-oriented communication systems (see, for example, \cite{huang2022semantic,Kalfa_2021,8792076}), which we review in more details in Section \ref{sec:related-work}. 

In this paper, we formulate a semantic-aware framework that is specifically tailored for \emph{object detection goal} in image transfer between the transmitter and receiver. 
To achieve this, we introduce two solution approaches to extract semantic information at the transmitter side. The first approach is based on expressing the objects (i.e., semantics) within an image using text, and then transferring text over wireless channel. At the receiver side, the receiver reconstructs the image and performs object detection. It should be noted that we assume that the object detection operation is performed on image datatype; therefore, if the receiver receives any other types of data, the received data should be converted to some form of image.  This approach is motivated by the scenarios in which the wireless channel is the limiting factor, i.e., limited data rates. The second approach to extract semantic information is based on the idea of removing irrelevant information (e.g., background) from images, and just transmitting detected objects (i.e., semantic) within an image.

Both of our system designs are inspired by auto-encoder networks, in which we have an encoder network/function that converts input to a feature vector. In this case, we convert images to intermediate data type (image or text), and the decoder function converts the intermediate data to the intended data, which in our case, is the objects within the original images. We implement both of these methods and test the efficacy of the proposed algorithms using open-source image datasets. Our numerical results demonstrate that the proposed system model and solutions provide performance gains in terms of the amount of data transmitted, without compromising the accuracy of the object detection goal. In summary, the main contributions of this paper are as follows:
\begin{itemize}
    \item We propose a framework for semantic-aware and goal-oriented communication for object detection in wireless end-to-end image transmission systems with limiting constraints. Such a model can be utilized in vehicular or satellite communication networks where ample computing resources are available at transmitter and receiver nodes, but due to the inherent unreliability and dynamic nature of the communication link, the data rate is low.
    \item We develop two end-to-end solutions to achieve object detection in image transfer systems. The first approach is based on representing the image semantics using text, and the second approach is based on extracting the semantics in terms of detected objects themselves, and removing irrelevant information such as image background. 
    \item We implement the proposed approaches, and test the efficacy of the solutions on open image datasets. Our implementation includes defining semantic extraction and reconstruction functions (i.e., encoder and decoder). For example, our numerical results show that by expressing semantic information using text, we can achieve 99\% reduction in data traffic, at the cost of slight increase in object detection error. 
\end{itemize}

\noindent 
The rest of this paper is organized as follows. In Section \ref{sec:related-work}, we review related works. In Section \ref{sec:system-model}, we present our proposed system model and problem formulation. Section \ref{sec:solution} includes our developed solution, followed by numerical results in Section \ref{sec:results}. Finally, Section \ref{sec:conclusion} concludes the paper.

 \section{Related Works}
 \label{sec:related-work}
\textbf{Semantic Communication Architectures.} 
The paper \cite{8792076} proposes a model-free approach that uses a reinforcement learning algorithm to train an end-to-end communication system to optimize a specific objective, such as maximizing the information transfer rate or minimizing the error rate. Yet, they did not consider the data and semantic information within data points. That is, their implementation objective is to achieve the best system considering the channel, not the semantic information.  The authors in \cite{Kalfa_2021} demonstrate a formal graph-based semantic language and a goal-filtering method that enables goal-oriented signal processing. The proposed semantic signal processing framework can easily be tailored for specific applications and goals in a diverse range of signal processing applications.
The authors in \cite{Sana2022learning} propose a Transformer-based semantic communication system architecture to learn the underlying semantics of the data being transmitted, communication systems can adapt to changing network conditions and optimize their performance. They evaluated the effectiveness of their approach in a natural language processing theme that supports their Transformer-based model. 

\textbf{Semantic-Aware Image Processing.}
The authors in ~\cite{huang2022semantic} propose a Reinforcement Learning-based Adaptive Semantic Coding (RL-ASC) approach to image semantic coding using deep learning techniques, with the goal of improving the efficiency and effectiveness of image communication. The approach includes a convolutional semantic encoder to extract semantic concepts, an RL-based semantic bit allocation model, and a Generative Adversarial Nets-based semantic decoder, which results in noise-robust and visually pleasant image reconstruction with reduced bit cost.

The authors in~\cite{7921920} presented an enhanced CNN model with the goal of image compression. Their model creates a map emphasizing important areas, making them better encoded compared to the background. They incorporated a comprehensive set of features for each class and applied a threshold to the total feature activations. This process results in a map highlighting semantically significant regions and improves their encoding quality compared to the background. 
Similarly, \cite{li2018learning} introduced a content-weighted image compression method using an importance map achieved by an importance map network, which utilizes the feature maps from the last residual block of the encoder as input. The importance map acts as a continuous alternative to discrete entropy estimation for compression rate control.
In both \cite{7921920} and \cite{li2018learning}, the error is measured in terms of Mean Square Error, and their objective is to find the best compression ratio for the exact reconstruction at the other end. 
In the work \cite{10100737}, they have established a transmission system for the images with limited bandwidth constraints. They used segmentation maps available with the COCO dataset for the semantic extraction block, and a pre-trained GAN network to reconstruct the images. Using this setup they could achieve a compression ratio of 20 percent. 

Compared with the existing works, we propose a novel approach to extract and express semantics of images for object detection goal. For instance, by expressing semantic features (i.e., detected objects) of an image using text, we can achieve a considerable gain (e.g., more than $90\%$) compared with traditional communication systems. 

\section{System Model and Problem Formulation}
\label{sec:system-model}
\subsection{System Model}
\label{subsec:h}
 
\textbf{Semantic-Aware Transmitter Model.} The proposed system model is shown in Figure \ref{fig:sysmodel}. In this model,
let us define domain \(\Omega_O\) as the original domain of data and domain \(\Omega_S\) as the semantic version of the first domain. We define the data in the original domain as \(X\) and function \(f_1()\) that transforms \(X\) to the semantic domain, i.e.: 
\begin{equation}
    f_1: \Omega_O \rightarrow \Omega_S.
\end{equation}
Then, we transmit the data \(f_1(X)\) over the wireless channel and receive  \(f_1(X)\). For object detection within images, the original domain is 3D matrix representation of images. In our solutions, the semantic domain could be either text or image.

\textbf{Semantic-Aware Receiver Model.}
At the receiver side, the opposite operation is performed. In particular, the receiver applies a function \(f_2()\) to the received data to transform it to the desired domain:
\begin{equation}
    f_2: \Omega_S \rightarrow \Omega_D.
\end{equation} 
The desired domain, \(\Omega_D\), can be the same as \(\Omega_O\) or any other domain based on the communication goal. 
Finally, we define \(Y\) as the reconstructed data, which is equal to \(f_2(f_1(X))\).  




\textbf{Semantic-Aware Performance Metrics.} The next step is to define a semantic evaluation function, \(h(.)\) to analyze \(X\) and \(Y\) in terms of semantic similarity. For instance, this function can be a general Mean Square Error function if \(\Omega_O\) and \(\Omega_D\) are the same and the goal is the exact reconstruction. In our system, the goal of the communication is an object detection task, then \(h(.)\) can be an object detector neural network that detects the objects in both \(X\) and \(Y\), and then we can simply calculate the distance between $h(X)$ and $h(Y)$. 
For example, if the image \(X\) that is being sent has three objects, \(O_1\), \(O_2\), and \(O_3\), where \(O_1\), \(O_2\), and \(O_3\) are the name of the classes in the object detection model, and we have \(n_1\)  instances of object (class) \(O_1\), and \(n_2\) and \(n_3\) for the objects \(O_2\) and \(O_3\), then \(h(X)\) would be the vector \([n_1, n_2, n_3]^T\).


\vspace{-.2cm}
\subsection{Semantic-Aware Object Detection Optimization}
With the aforementioned components, next we define our problem in terms of defining the $f_1()$ and $f_2()$ functions such that we achieve two objectives: \textbf{(i)} \textbf{semantic error} between \(X\) and \(Y\) is small, meaning that the \emph{goal} of communication is accomplished with high accuracy, and \textbf{(ii)} semantic transmissions provide \textbf{communication gains} by removing as much irrelevant data as possible. We define these metrics as follows: 
\begin{itemize}
    \item \textbf{Semantic Error.} This metric measures the occurred error in the reconstructed data with respect to the original data: 
        \begin{equation}\label{Rerror}
     E (X, Y) = \frac{\|h(X ) - h(Y) \|}{ \|h(X)\| }.
 \end{equation}

In the previous example, if the receiver detects the objects  \(O_1\), \(O_2\), and \(O_3\) but the number of detected instances is different and equal to \(\Acute{n_1}\), \(\Acute{n_2}\), and \(\Acute{n_3}\), \(h(Y)\) would be the vector \([\Acute{n_1}, \Acute{n_2}, \Acute{n_3}]^T\) and error is equal to the norm of distance between \(h(X)\) and \(h(Y)\).

\item  \textbf{Communication Gain.} We define communication gain as the amount of data transfer savings achieved by transmitting \(f_1(X)\) instead of original data $X$, i.e.,:
\begin{equation}\label{gain}
    G(X , f_1(X)) = 1- \frac{\mathcal{S}(f_1(X))}{\mathcal{S}(X)},
\end{equation}
where \(\mathcal{S}()\) is the size function which measures the binary size of data. For instance, the size of $M\times M$ 3D images would be \(M \times M \times 3 \times 64\) in a 64-bit system. 
\end{itemize}
By combing Eq. \eqref{Rerror} and Eq.~\eqref{gain}, we introduce a new metric as \emph{weighted error} that captures the balance between errors and gains. This metric multiplies the gain factor as a weight to the error, i.e.,:
    \begin{equation}\label{Wgain}
    E_w =  \Bigl[1 - G\bigl(X,f_1(X)\bigr)\Bigr] E (X, Y).
    \end{equation}
The weighted error metric implies that the higher gain simply is not sufficient. Instead, it should guarantee some level of accuracy. For example, if no data is transmitted, then the gain would be about $100\%$, but the accuracy of the object detection system will be zero. Also, in traditional communication, the gain is zero, and achieving low error might not be suitable with system constraints. 





Given these performance metrics, we aim to choose the mapping functions $f_1$ and $f_2$ such that the Weighted Error metric is minimized. However, in practice, there are several constraints on the semantic communication system, including data rate and error limit. 
In particular, suppose the wireless channel is a bottleneck for our problem, such that the data rate should be less than \(R_0\). As a result, the below constraints will be added to the optimization problem:
\begin{equation}\label{r0_cond}
    \mathcal{S}(f_1(X)) < R_0, 
\end{equation}
which can be translated to the below constrain based on the semantic gain: \( g(X, f_1(X)) > g_0\), where \(g_0 = 1 - \frac{R_0}{\mathcal{S}(X)}\).
On the other hand, there may be a condition on the acceptable minimum accuracy or maximum error; meaning that the error should be less than \(\epsilon_0\) (\(E_{r_{f_1,f_2}} < \epsilon_0 \)). 
Therefore, we have:

 \begin{equation} \label{eq:formulationNew1}
\begin{aligned}
\min_{f_1, f_2} \quad & \Bigl[1 - G\bigl(X,f_1(X)\bigr)\Bigr] E (X, Y) \\
\textrm{s.t.} \quad & \mathds{E} \Bigl[G(X, f_1(X))\Bigr] > g_0,  \\
  & \mathds{E} \Bigl[ E (X, Y) \Bigr]<\epsilon_0, 
\end{aligned}
\end{equation}
where $\mathds{E} [.]$ denotes the expected value. Next, we introduce two methods for defining  $f_1$ and $f_2$. 




\section{Semantic-Aware Object Detection}
\label{sec:solution}
With the goal of object detection in mind, we present two solutions for extracting semantic information at the transmitter side. The first approach is based on using text to express image semantics (i.e., objects) and then transferring text over a wireless channel. The receiver reconstructs the image and detects objects on the receiving end. This solution approach is motivated by scenarios in which a wireless channel with low data rate is the limiting factor. We refer to this scenario as ``Limited Data Rate System''. 
The second method of extracting semantic information is based on the concept of removing irrelevant information (e.g., background) from images and only transmitting detected objects (i.e., semantic) within an image. 
This is referred to as ``Low Error Tolerance System''. Both of these approaches are built upon the idea of auto-encoder systems, as shown in Fig. \ref{fig:sysmodel}. 

\begin{algorithm}[t]
\caption{Low Data Rate System}
\SetKwInOut{Input}{Input}
\SetKwInOut{Output}{Output}
\Input{A set of images $images$}
\Output{Error and gain values for each image}
\ForEach{image $X_i$ in $images$}{
  \textbf{Step 1:} Convert image $X_i$ to a list of 5 descriptive texts $T$ using a neural network $f_1(X_i)$\;
  \textbf{Step 2:} Send $T$ over the communication channel\;
  \textbf{Step 3:} Initialize $\text{Error}(X_i)$ to 0\;
  \ForEach{caption $T_n$ in $T$}{
    Generate an image $R_{X_i}$ based on $T_n$ using neural network $f_2(T_n)$\;
    Calculate the semantic representation for $R_{X_i}$ as $h(R_{X_i})$\;
    Add $h(R_{X_i})$ to $h(Y_i)$\;
  }
  $h(Y_i) =  \frac{1}{length(T)} h(Y_i)$\;
  Calculate $\text{Error}(X_i, Y_i) =  ||h(X_i) - h(Y_i)||$\;
  Calculate $\text{Gain}(X_i) = 1 - \sum_{n=1}^{5} \frac{\mathcal{S}(f_2(T_n))}{\mathcal{S}(X_i)}$\;
  \textbf{Output:} $\text{Error}(X_i, Y_i)$ and $\text{Gain}(X_i)$ for image $X_i$\;
}
\label{alg:dataRateSolver}
\end{algorithm}

\noindent 
\textbf{Low Data Rate System.}
First, assume that we have a low channel data rate, and the receiver has a higher tolerance for error; meaning that the receiver can accept some level of error but needs the data immediately, and under the low channel data rate, sending data with traditional communication methods does not satisfy the receiver's requirement of freshness. Therefore, one approach is to extract semantic information and express it with as little data as possible, so that with a given low channel rate, the transmission time is as short as possible. To this end, we choose function \(f_1()\) to be an \emph{image-captioning function}, meaning that at the transmitter side, \(f_1()\) converts an image to some text describing the image, and the data that is being sent to the receiver is the text describing the image. At the receiver side, we have the function \(f_2()\) that converts the text to an image. In this scenario, the original and desired domains are 3D matrix domains that are fixed and defined by the goal of the communication, but with this specific choice of the function \(f_1()\), the semantic domain is text-domain. 

Algorithm~\ref{alg:dataRateSolver} describes the steps to implement this method, which takes a collection of images as input and aims to evaluate the quality of both textual descriptions and images generated from these input captions. This algorithm quantifies the error between the generated image and the original image by measuring their semantic representations through the function \(h()\), utilizing the Euclidean distance for the object detection goal. 

\noindent 
\textbf{Low Error Tolerance System.}
The second model is when the receiver has a lower tolerance for error but the channel rate is higher but still not enough for performing a traditional method of communication. In this case, we chose the function \(f_1()\) as the object extractor function which basically, detects the objects within the given images and returns the objects in those images. The semantic domain will be the 3D\(\times\)$N$ matrix domain where $N$ is the number of detected objects. The function \(f_1()\) removes the unnecessary information such as background from images, and returns the semantics of data which are the objects. The function \(f_2()\) for this example does not need to be implemented since the receiver can understand the output of function \(f_1()\).  
Algorithm~\ref{alg:limitedError} describes the steps for implementing object detection for the limited error tolerance scenario. It takes a set of images as input and aims to evaluate the quality of object detection and image understanding. The algorithm processes each image, converting it into a list of its constituent objects using a neural network. These object representations are transmitted over a communication channel. For each object, a semantic representation is computed using the function \(h()\) and accumulated into a list. The error is then calculated by assessing the mismatch between the semantic representation of the entire image and the list of object semantic representations. 
The algorithm calculates the communication gain and semantic error values for each image. 

\begin{algorithm}[t]
\caption{Low Error Tolerance System}
\SetKwInOut{Input}{Input}
\SetKwInOut{Output}{Output}
\Input{A set of images $images$}
\Output{Error and gain values for each image}
\ForEach{image $X_i$ in $images$}{
  \textbf{Step 1:} Convert image $X_i$ to a list of its objects $O_{X_i}$ using a neural network $f_1(X_i)$\;
  \textbf{Step 2:} Send $O_{X_i}$ over the communication channel\;
  \textbf{Step 3:} Initialize $\text{Error}(X_i)$ to 0\;
  Initialize an empty list $d_{O_{X_i}}$\;
  \ForEach{object $o$ in $O_{X_i}$}{
    Compute $d_{o} = h(o)$ where $h()$ is an object detection function\;
    Add $d_{o}$ to $d_{O_{X_i}}$\;
  }
  $h(Y_i) = d_{O_{X_i}} $\;
  Calculate $\text{Error}(X_i, Y_i) = ||h(X_i) - h(Y_i)||$\;
  Calculate $\text{Gain}(X_i) = 1 - \frac{\mathcal{S}(d_{O_{X_i}})}{\mathcal{S}(X_i)}$\;
  \textbf{Output:} $\text{Error}(X_i, Y_i)$ and $\text{Gain}(X_i)$ for image $X_i$\;
}
\label{alg:limitedError}
\end{algorithm}
\begin{figure}[t]
  \begin{subfigure}[b]{\columnwidth}
     \includegraphics[width=\textwidth]{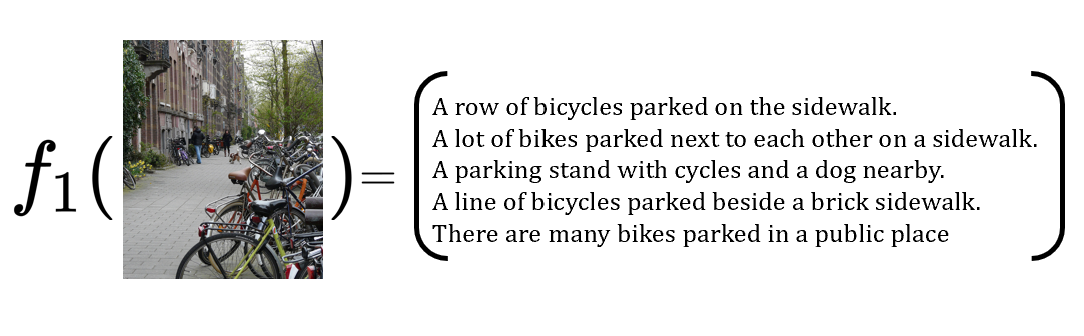}
    \caption{{Semantic-aware transmitter for object detection}}
    \label{fig:f11}
  \end{subfigure}
  \hfill 
  \begin{subfigure}[b]{\columnwidth}
    \includegraphics[width=\textwidth]{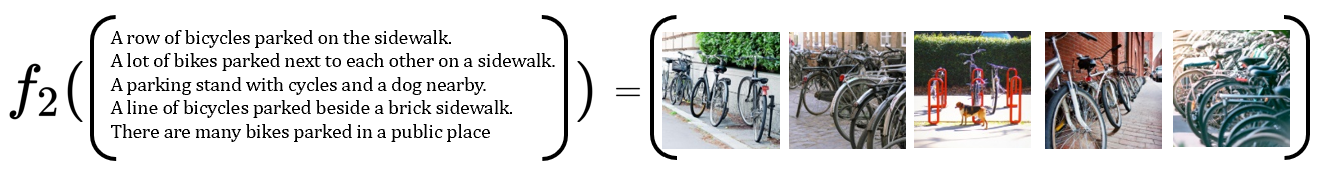}
    \vspace{-.5cm}
    \caption{{Semantic-aware receiver for object detection}}
    \label{fig:f21}
  \end{subfigure}
  \caption{{(a) Semantic transmitter operation. The  input image is described using $5$ different captions, which are generated by image-captioning neural network. (b) Semantic receiver operation. Received captions/texts are translated into an image. The goal of communication is to accurately detect the objects in the original image. } }
\end{figure}

\section{Simulation Results}
\label{sec:results}
In this section, we examine the efficacy of the proposed solutions  using open source image datasets. As described, the goal of our semantic-aware system is object detection in images. Hence,  
in all simulation cases, the original domain is the image domain, and based on the communication goal and constraints we decide on functions \(f_1()\) and \(f_2()\), \(h()\) and semantic domain \(\Omega_S\). 
For each setup, we plot the error performance as well as the weighted error performance for semantic vs. traditional communication. All the reported results are \emph{cumulative average performance}, as more images are being included in the averaging process. 



\subsection{Simulation Setup}
\textbf{Dataset Description.} The dataset we primarily use is the COCO Validation Image 2014 dataset \cite{data-set}. This subset is designed to test the performance of computer vision algorithms, encompassing object detection, image segmentation, and captioning. It comprises thousands of hand-annotated images, with varying content, scene complexity, and object categories. Each image in this subset is accompanied by detailed annotations, specifying object categories, object locations, and textual descriptions. The dataset spans a wide range of object categories, from people and animals to vehicles and household items.

\textbf{Object Detection Module.} We utilize  the \emph{RetinaNet Resnet50 FPN} as the object detector in the system. The object detector model is a customized version of the RetinaNet object detection framework. It uses a ResNet-50 backbone and incorporates a Feature Pyramid Network (FPN). This specific model has been trained on the widely recognized COCO (Common Objects in Context) dataset.
The RetinaNet model, initially introduced in the paper \cite{coco-object-detector-model}, is renowned for its ability to efficiently and accurately identify objects in images.

\textbf{Image Captioning and Text-to-Image Modules.} 
In the first scenario, we express semantic information of images using text. Therefore, we needed to automate the process of image captioning as well as image generation from text. 
For the image-generator function, we utilized the DALL-E2 API to generate images for the given descriptive text. DALL-E2 is an extension or a new version of OpenAI's DALL-E, which is known for generating images from textual descriptions. DALL-E is based on a GPT-3-style architecture, which combines a transformer network with a VQ-VAE-2 (Vector Quantized Variational Autoencoder 2) architecture. The transformer network is responsible for understanding textual input, and the VQ-VAE-2 generates the corresponding image. The original DALL-E model has hundreds of millions of parameters, which are distributed across multiple layers. Furthermore, we utilized COCO API for caption generation. For each image, we generated five captions describing the image and sent a list containing all the captions over the communication channel. On the receiver side, we used DALL-E2 API to generate one image for each received caption. It should be noted that any numbers of captions can be generated, and we have arbitrarily chosen $5$ to balance the trade-off between running time and performance.



\begin{figure}[t]
     \centering
     \begin{subfigure}[b]{.485\linewidth}
         \centering
         \includegraphics[width=\linewidth]{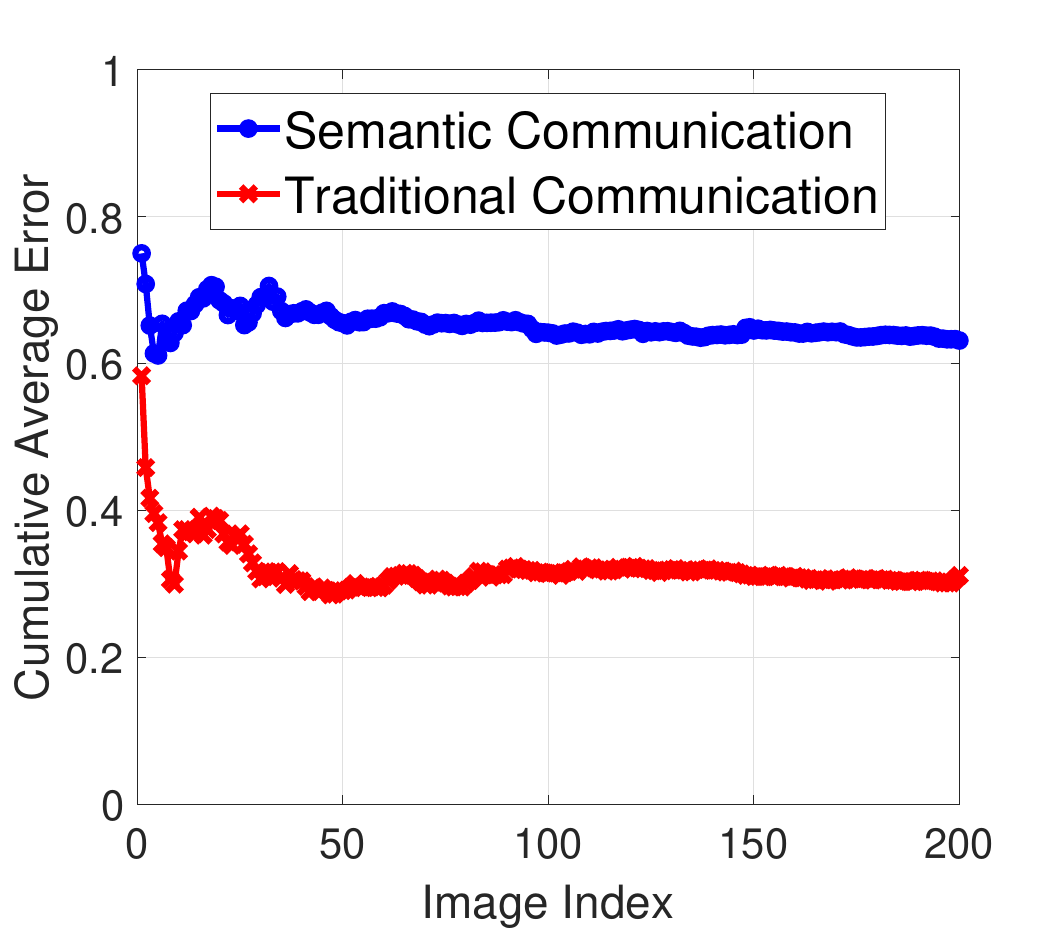}
         \caption{Error Performance}
         \label{fig:Error200-sen1}
     \end{subfigure}
     \begin{subfigure}[b]{.49\linewidth}
         \centering
         \includegraphics[width=\linewidth]{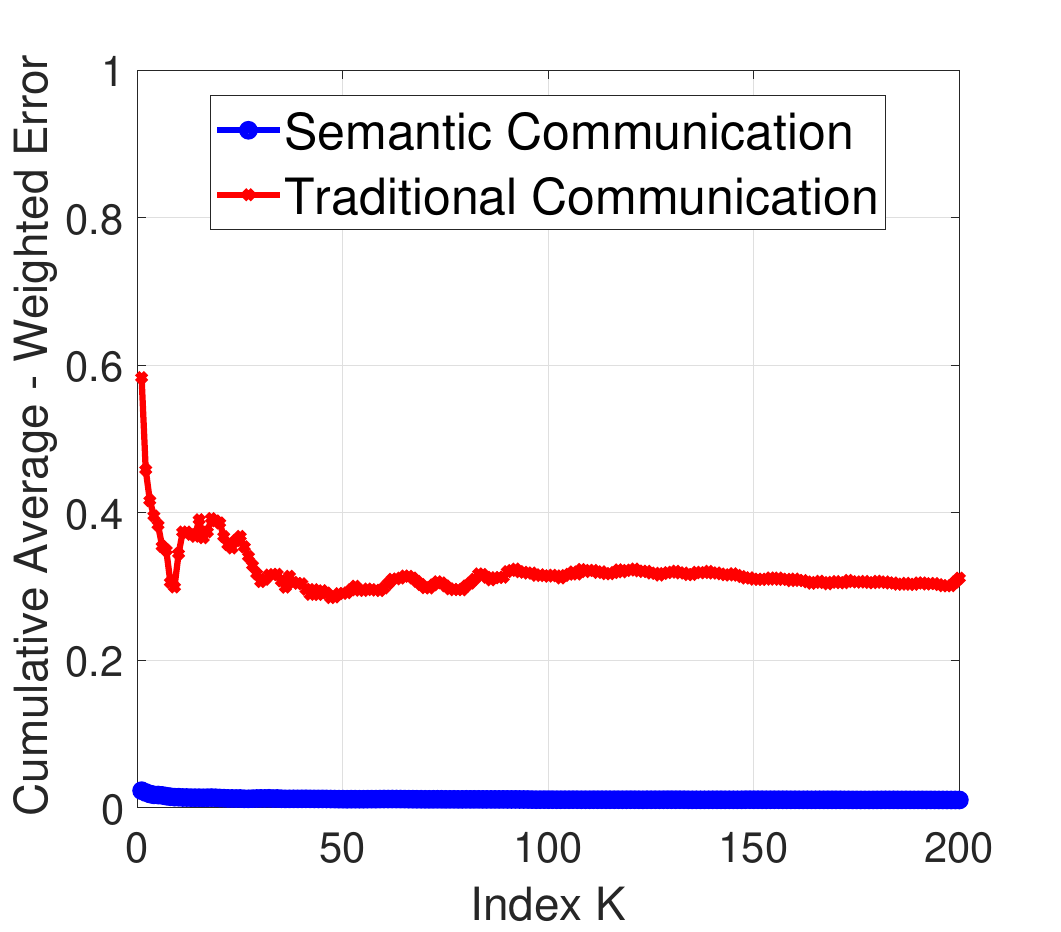}
         \caption{Weighted Error Performance}
         \label{fig:WgainGain-sen1}
     \end{subfigure}
        \caption{Simulation results for the scenario in which text is used to describe the semantics images.}
        \label{fig:cp-1-Sim}
\end{figure}
\subsection{Results for Low Data Rate System}\label{low-data-rate-system}
In this case, there is a limit over the data rate that the channel can handle; therefore, we want to reduce the size of transmitting data as much as possible. Transforming images to a segment of text that describes the objects in the images and their environment is one of the options that would result in a high gain. Therefore, our algorithm processes each image and uses the COCO caption generator neural network to convert it into a list of $5$ descriptive texts (see, for example, Fig.~\ref{fig:f11}). These texts are then transmitted over a communication channel. For each descriptive text, another neural network, Dall-E2, generates an image (see Fig. \ref{fig:f21}), and the algorithm quantifies the semantic error between this generated image and the original image. The error is computed as the Euclidean distance between semantic representations which is measured by the function \(h()\) of the two images. The algorithm also calculates the gain value, reflecting data rate efficiency, and an average error to assess the quality of the generation process.

Given this configuration, we set  \(\epsilon_0 = 0.65\) and \(g_0 = 0.9\),  obtain the error and weighted error performance as shown in Figure \ref{fig:cp-1-Sim}. As shown in Figure \ref{fig:WgainGain-sen1}, since we converted images to text, we could achieve a considerable amount of gain over all the data points. The average gain for this setup is 99\(\%\), and while the average error is 65\(\%\) satisfying the error constraint, the objective value of the weighted error is 1\(\%\).
Here, the key factor is the receiver's ability to comprehend the received text (Natural Language) and convert it to an image. We can improve the accuracy by generating more descriptive texts at the transmitter side and training a more powerful machine learning model at the receiver side for image generation.

\begin{figure}[t]
     \centering
     \begin{subfigure}[b]{0.49\linewidth}
         \centering
         \includegraphics[width=\linewidth]{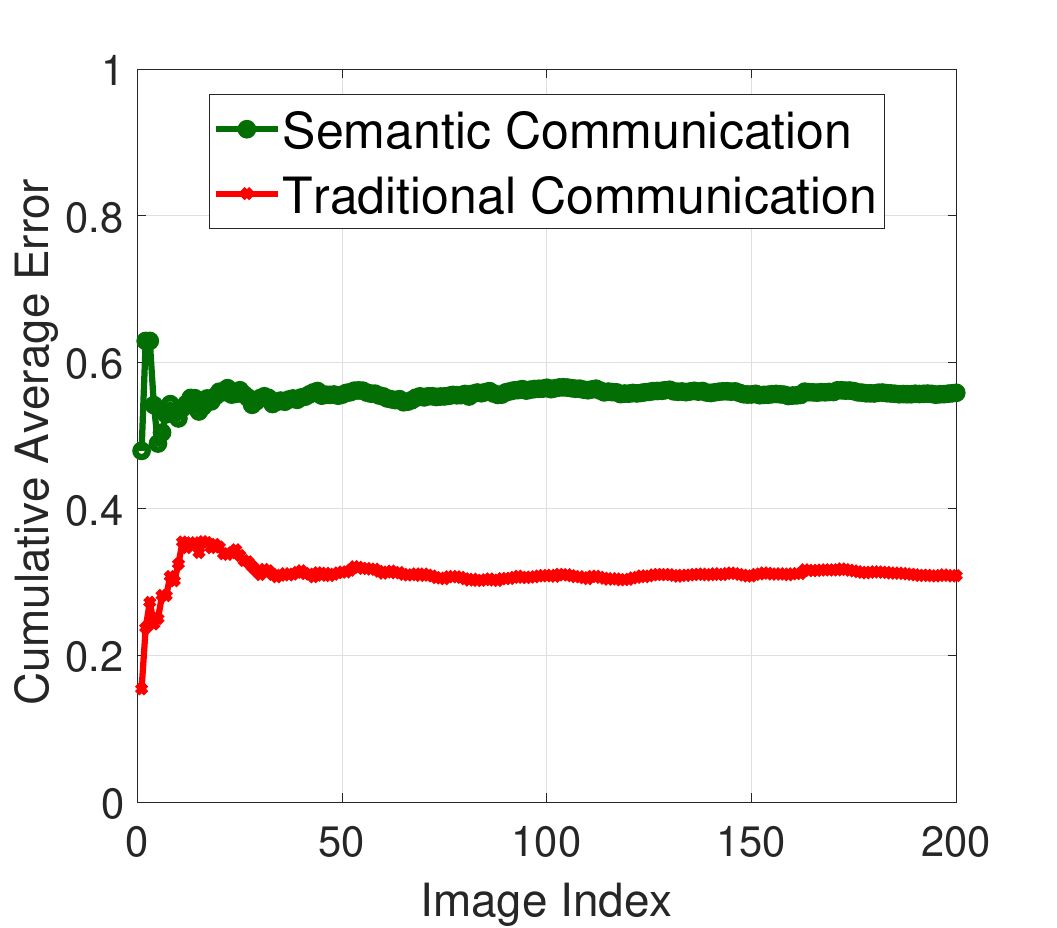}
         \caption{Error performance}
         \label{fig:Error200-sen2}
     \end{subfigure}
     \begin{subfigure}[b]{0.49\linewidth}
         \centering
         \includegraphics[width=\linewidth]{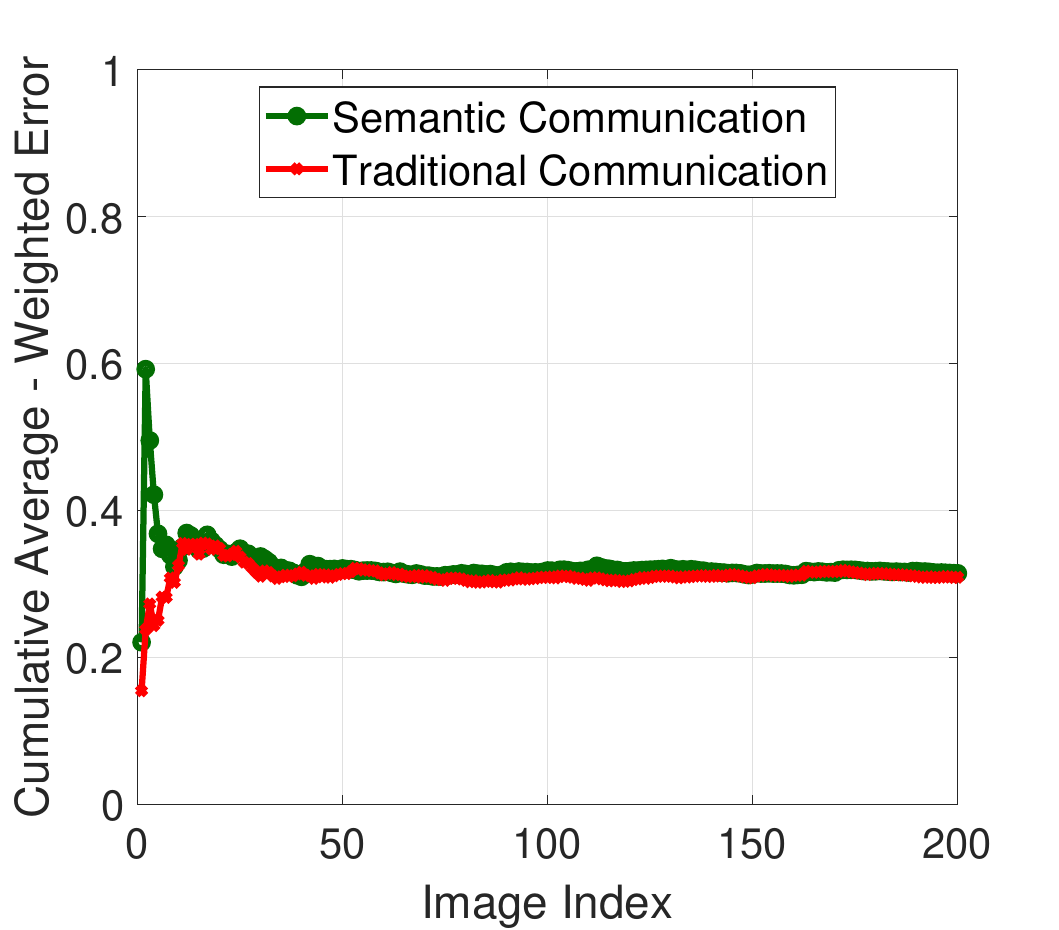}
         \caption{Weighted Error Performance}
         \label{fig:WgainGain-sen2}
     \end{subfigure}
      
        \caption{Simulation results for the scenario in which the transmitter extracts objects (semantics) and removes irrelevant information (such as background) from images.}
        \label{fig:cp2-results}
\end{figure}

\subsection{Results for Low Error Tolerance System}\label{low-error-tolerance-system}
For this configuration, the transmitter sends the detected objects in the images. 
In order to implement this system,  we used the COCO object detection pre-trained model. We extracted the objects in the images and sent objects (without the background) over the channel. The receiver, therefore, just utilizes the extracted objects. Since the goal is to reconstruct objects, we define  
the  semantic error \(E(X, Y)\) as the mismatch between outputs of \(h()\) function when applying on \(X\) and each \(Y\).
The \(f_2()\) function in this case is an identity function.

To assess the system performance, we implemented Algorithm~\ref{alg:limitedError}, which takes a set of images as input and focuses on evaluating the quality of object detection and image understanding. The algorithm iterates over each image, utilizing a neural network, denoted as \(f_1(X_i)\), to convert the image into a list of its constituent objects, represented as \(O_{X_i}\). These object representations are then transmitted over a communication channel. For each object in \(O_{X_i}\), the algorithm computes a semantic representation using an object detection function \(h(Y_j\) and accumulates these representations into a list \(d_{O_{X_i}}\). The error for the image \(X_i\) is calculated by assessing the mismatch between the semantic representation of the entire image \(h(X_i)\) and the semantic representation list \(d_{O_{X_i}}\). The algorithm also calculates the gain to evaluate data rate efficiency, which is expressed as the difference between 1 and the ratio of the size of the semantic representation list \(d_{O_{X_i}}\) to the size of the original image \(X_i\). 
The results, including error and gain values, are then output for each image in the dataset. 
In this system, we set \(\epsilon_0 = 0.55\) and \(g_0 = 0.5 \), resulting in an average gain of 50\(\%\), an average error of 55\(\%\), and an average weighted error of 30\(\%\).  From the results reported in Fig.~(\ref{fig:cp2-results}), we observe that although the error is in an acceptable range, the combination of error and gain has a lower range. In other words, having a very high gain while the error is very high as well, is meaningless and contradicts the point of communication. 




\section{Conclusion}
\label{sec:conclusion}
In this paper, we presented a semantic-aware framework tailored for object detection in the transfer of image datasets between a sender and a receiver. To accomplish this, we proposed two methods for extracting semantic information at the sender's end. The first method involves describing objects in an image using text and transmitting this text over a wireless channel. At the receiver's end, the image is reconstructed, and object detection is performed. The second method for extracting semantic information focused on eliminating irrelevant data (e.g., background) from images and transmitting only the detected objects (i.e., semantic content) within the image. We proposed and implemented the algorithms to achieve these solution approaches. Our numerical results using open source image datasets show that these approaches are effective, implying that semantic-aware methods have the potential to be a useful tool for optimizing communication systems that are tailored to achieve specific goals.

\section*{Acknowledgment}
The material is based upon work supported by NSF grants 1955561, 2212565, and 2323189. Any opinions, findings, conclusions, or recommendations expressed in this material are those of the author(s) and do not necessarily reflect the views of NSF.

\bibliographystyle{IEEEtran}
\bibliography{ref}

\end{document}